\documentclass[fleqn,usenatbib]{mnras}

\usepackage{newtxtext,newtxmath}

\usepackage[T1]{fontenc}

\DeclareRobustCommand{\VAN}[3]{#2}
\let\VANthebibliography\thebibliography
\def\thebibliography{\DeclareRobustCommand{\VAN}[3]{##3}\VANthebibliography}

\newcommand{\ixpe}{\textit{IXPE}}

\usepackage{graphicx}	
\usepackage{amsmath}	
\usepackage{bm}	

\title[X-ray polarization from an ADC source]{The detection of high X-ray polarization from an accretion disc corona source and its modelling via Monte Carlo radiation transfer simulation}

\author[R. Tomaru, C. Done, H. Odaka]{
Ryota Tomaru$^{1}$\thanks{E-mail: r.tomaru.sci@osaka-u.ac.jp}
Chris Done$^{2,3}$
Hirokazu Odaka$^{1}$
\\
$^{1}$Department of Earth and Space Science, Graduate School of Science, Osaka University, 1-1 Machikaneyama, Toyonaka, Osaka 560-0043, Japan\\
$^{2}$Centre for Extragalactic Astronomy, Department of Physics, Durham University, South Road, Durham DH1 3LE, UK\\
$^{3}$Kavli Institute for Physics and Mathematics of the Universe (WPI), University of Tokyo, Kashiwa, Chiba 277-8583, Japan
}

\date{Accepted XXX. Received YYY; in original form ZZZ}

\pubyear{2023}

\begin{document}
\label{firstpage}
\pagerange{\pageref{firstpage}--\pageref{lastpage}}
\maketitle

\begin{abstract}

We report a time averaged 2-8~keV X-ray polarization degree (PD) of $8.5 \pm 1.6\%$ ($>3\sigma$ detection) from the accretion-disc–corona (ADC) neutron-star system 2S 0921–630 (=V395 Car) observed with the Imaging X-ray Polarimetry Explorer ({\ixpe}).
As the observation includes an eclipse, we analyze eclipse and out-of-eclipse intervals separately. 
The eclipse PD is $15\pm3\%$, compared to $5.9\pm1.9\%$ out of eclipse, with no clear evidence for an associated change in polarization position angle (PA).
We also search time averaged, eclipse and non-eclipse spectra and find marginal evidence ($2\sigma$) for a change in PA with energy, and even weaker evidence for an increase in PD with energy. 
We use a Monte-Carlo spectropolarimetric radiation transfer simulation to model the polarization produced from a disc accreting neutron star, combining boundary-layer emission, its disc reflection, and the disc continuum, each with its intrinsic polarization. The model then also includes scattering of this composite spectrum in the column density distribution produced by a thermal–radiative wind launched by X-ray irradiation of the outer disc. 
At high inclination angles, where the observed flux is seen only via scattering in the wind, this model can reproduce both the observed PD and its (very weakly significant) increase with energy.
However, it does not predict the stronger (but still only marginally significant) change in PA with energy.
If this is a real effect then it points to a more complex, non-axisymmetric scattering geometry than that assumed in our model. 

\end{abstract}




\section{Introduction}\label{sec:intro}
Determining the geometry of accretion flows in low-mass X-ray binaries (LMXBs) is challenging when relying on spectroscopy and timing alone, because multiple different configurations can reproduce similar spectra and variability.
Instead, X-ray polarimetry directly probes geometry as it is imprinted by departures from spherical symmetry. The polarization degree (PD) and polarization angle (PA) carry the history of how the emitted radiation is scattered and reprocessed in non-spherical media \citep{Chandrasekhar1960}.
With the Imaging X-ray Polarimetry Explorer ({\ixpe}), energy-resolved polarimetry of Galactic compact objects has become feasible, enabling geometric tests of models that are otherwise degenerate \citep{Weisskopf2022}. 
Early {\ixpe} observations of "typical" weakly magnetized neutron-star LMXBs reported low PDs ($\sim 1-2\%$), consistent with a largely direct view of the central emitter and only modest scattering (see e.g. the review by \citealt{Ursini2024}).

Accretion-disc–corona (ADC) systems provide a contrasting, scattering-dominated regime.
These are systems viewed at such high inclination that the vertically extended outer disc obscures the central engine. The observed continuum is produced predominantly by electron scattering and reprocessing in a photoionized medium above the outer disc, plausibly an equatorial thermal–radiative wind \citep{White1982b, Jimenez-Garate2002, Kallman2003, Church2005, Ponti2012, Tomaru2023b}.
Such a scattering-dominated configuration naturally predicts substantially higher PD than in direct-view systems, along with a diagnostic energy dependence caused by the relative weighting of disc-like versus more isotropic illumination components \citep{Tomaru2024}.

2S 0921–630 (=V395 Car) is a prototypical ADC source. 
This source shows regular
eclipses in the X-rays, which gives the orbital period of 9 days and requires a very high inclination \citep[$i\sim82^{\circ}$]{Ashcraft2012}. 
The companion star is an evolved K0III (subgiant) in orbit around a neutron star \citep{Shahbaz2007,Steeghs2007, Ashcraft2012}, with a Gaia DR3 distance of $9.2^{+2}_{-1.4}$~kpc \citep{Avakyan2023} with Baysian estimation \citep{Bailer-Jones2018}.
The observed X-rays have low apparent luminosity, 
and there are multiple emission lines, 
both of which indicate that the intrinsic X-ray flux is geometrically suppressed, so that reprocessing in the wind dominates the spectrum \citep{Shahbaz2007,Ashcraft2012, Tomaru2023b}
These properties make this object an ideal laboratory for testing polarimetric predictions for scattering-dominated geometries.

Here we present the first X-ray spectro-polarimetry data of 2S 0921-630. Our data also includes an eclipse, making this the second eclipsing ADC seen by {\ixpe} \citep{Mikusincova2023}.
We report a high time-averaged polarization of $PD = 8.5 \pm 1.6\%$ in 2–8 keV band. There are hints of an
increase in PD to $15\pm 3$\% during the eclipse, and hints of a swing in PA and PD with energy (Section 2)
but none of these are formally 
significant ($<3\sigma$: Section 2). Hence we use the time averaged data and construct a model to interpret the spectro-polarimetry via Monte Carlo radiation transfer  (Section 3). We show that the time averaged PD is well matched to the expectations of wind scattering (Section 4). The model also predicts an increase in PD with energy consistent with the data, but does not predict any change in PA with energy (Section 4). We briefly discuss the effect of 
additional non-axisymmetric structures which may produce a PA swing (Section 4), 
but the data are not good enough to support more detailed modelling. We summerize our results in 
Section 5.

\section{Data analysis}

\subsection{Obseved data}
\begin{figure}
    \centering
    \includegraphics[width=0.9\linewidth]{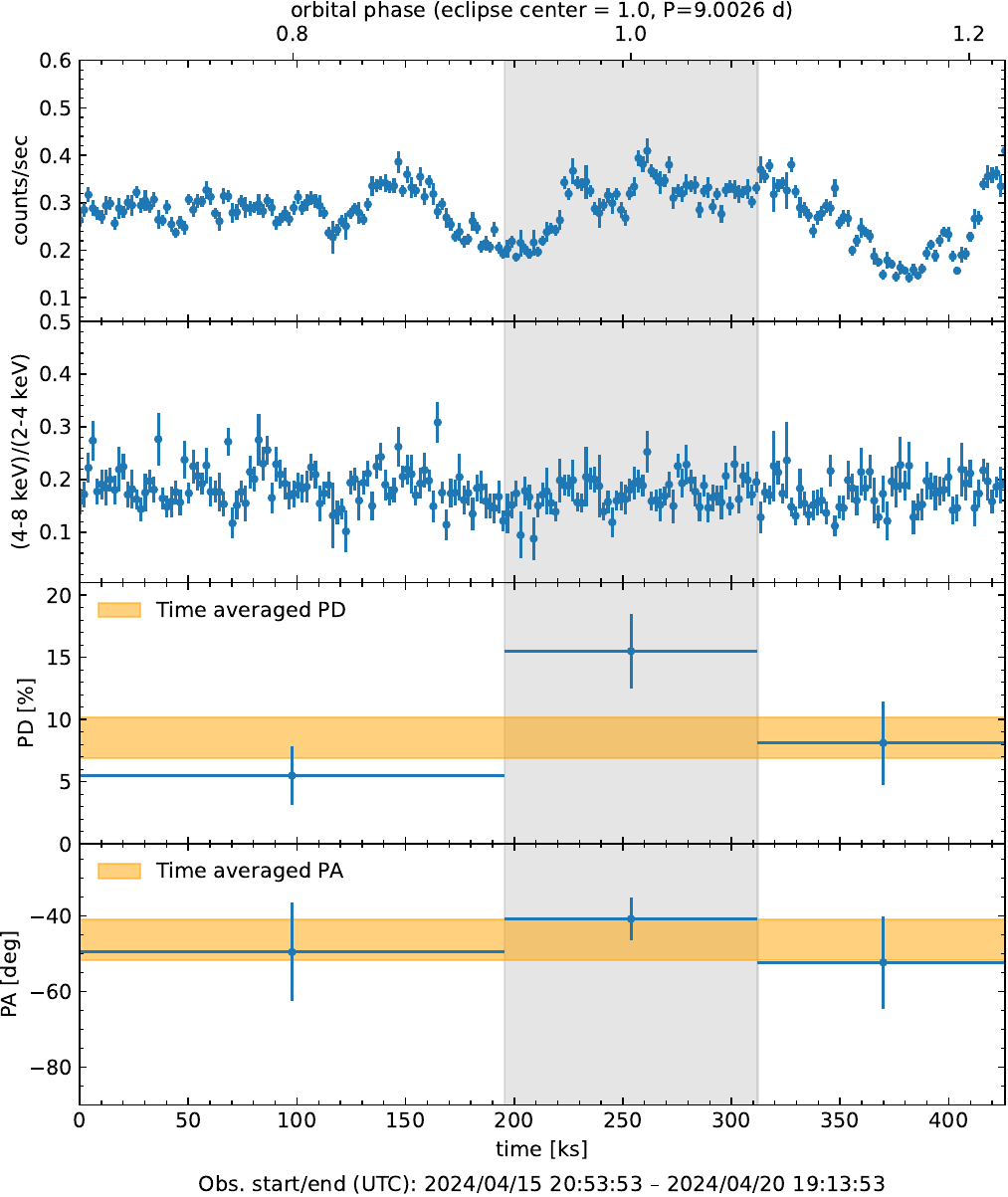}
    \caption{The {\ixpe} light curve of 2S 0921–630 with 1000s time binning. The errors in PD and PA show 1 $\sigma$ confidence level.
    The orange-shaded region shows the time average PD and PA. 
    The grey-shaded region shows the predicted eclipse phase (0.925-1.075) based on the optical and infrared observations \citep{Ashcraft2012}.
    }
    \label{fig:lc}
\end{figure}
The {\ixpe } \citep{Weisskopf2022} observed 2S0921 with observation ID 03001201 from 20:55:02 UTC on 15 April 2024 to 19:43:45 UTC on 20 April 2024. 
We downloaded Level 2 processed data, 
suitable for scientific data analysis, from the {\ixpe} HEASARC archive.
The source region and the background region were spatially selected within the {\ixpe} field of view, defining concentric areas centred on the image centroid. 
The source region is a circular area with a radius of 1.0 arcminutes for each of the three detector units (DUs), while the background region is the same radius but a different region without a source. 

The source intensity, binned on 1000s,
shows only modest (factor 2) variability during the observation
(top panel, Fig. \ref{fig:lc}).
The hardness ratio, defined as the ratio of counts in 4-8 keV to 2-4 keV, does not show any strong departures from a constant, showing that there was no strong spectral variability during the observation, such as associated with state transitions (2nd panel in Fig.\ref{fig:lc}).

We use the 
orbital ephemeris of \citet{Ashcraft2012} to derive the orbital phase. This gives a prediction for the 
eclipse centre date $T_0(HJD) = 2453397.28(2) + 9.0026(1)E$ \citep{Ashcraft2012}, where $E$ is the integer cycle count and its width is about $0.15$ in phase. Our data then cover 
orbital phase 0.67-1.22, and include the eclipse (grey shaded region, Fig.\ref{fig:lc}). 
The X-ray light curve of {\ixpe} does not show a clear eclipse signature, but partial eclipses are expected due to the 
scattering-dominated nature of ADC sources \citep{Mason1987}. Nonetheless, there are broad dips in the lightcurve though these appear slightly before and after the predicted dip ephemeris.
We extract the time averaged spectro-polarimetric data for the whole observation (phase 0.67--1.22), as well as separately for the eclipse (phase 0.925--1.075) and non-eclipse (phase 0.67-0.925 and 1.075-1.22) intervals.

Following the Quick Start Data Analysis Guide, we derive the spectropolarimetric data from the Level 2 data using {\sc xselect}. 
We chose the statistic weight "NEFF" and created Ancillary Response Files (ARFs) and Modulation Response Files (MRFs) using {\it ixpecalcarf} with an extraction region of 1 arc min radius.
We also group the data to ensure that each energy bin contains at least 20 counts for each DU for Stokes I data using {\sc grppha}, to allow the use of $\chi^2$ statistics in spectral fitting, and group the Stokes Q and U data to match the I data bins.

\subsection{Spectro-polarimetric analysis}

\begin{figure}
    \centering
    \includegraphics[width=0.9\hsize]{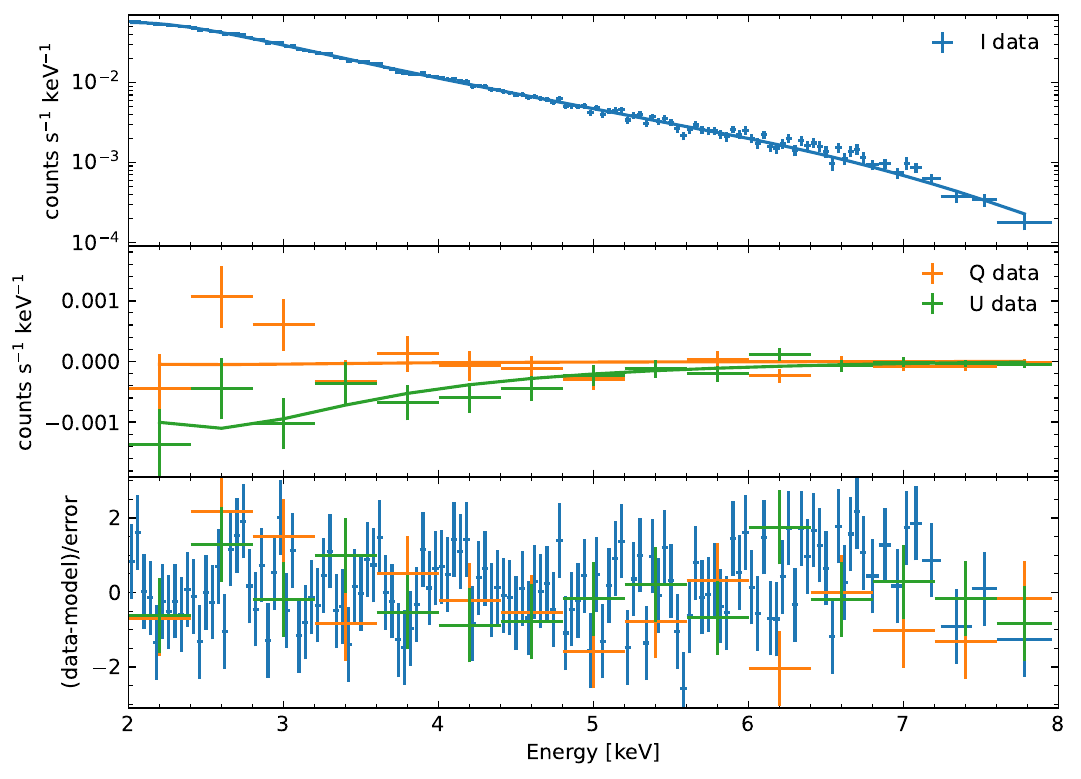}
    \caption{The time-averaged spectrum (blue), Stokes Q (orange), and Stokes U (green).
    The bottom panel shows the residuals of the fit with the model, which is consistent with the data in $2\sigma$. 
    The data are grouped into 3 DU data, and binned for plotting purposes.
    }
    \label{fig:sp_pd}
\end{figure}

\begin{figure*}
    \centering
    \includegraphics[width=0.9\linewidth]{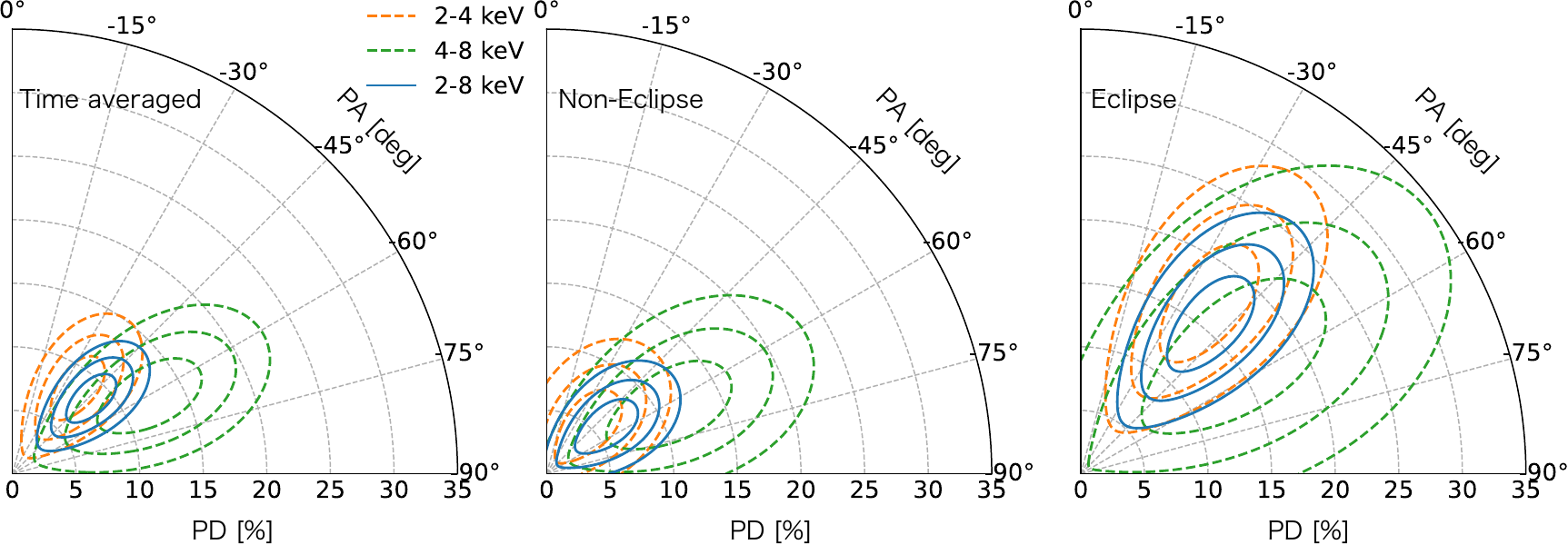}
    \caption{The polarization 2D contours taken by spectro-polarimetric fitting for each time intervals(Tab.\ref{tab:time_average}).
    The contour lines show confidence levels of $1\sigma  =68.3\%$, $2\sigma =95.5\%$, and $3\sigma =99.7\%$. }
    \label{fig:pol_whole}
\end{figure*}

\begin{figure}
    \centering
    \includegraphics[width=0.9\hsize]{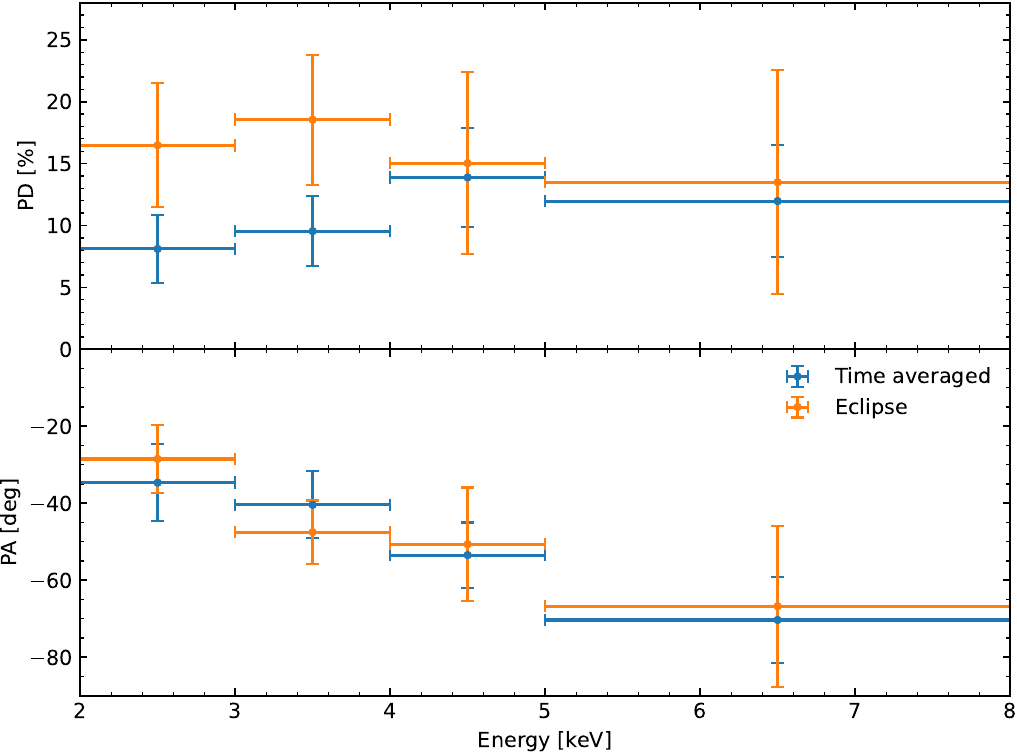}
    \caption{The energy dependence of PD and PA taken by spectro-polarimetric fitting. The blue points show the time averaged data, while the orange points show those from the eclipse phase. The 2-4 keV band of the eclipse phase clearly shows the higher PD than the time-averaged, whereas the PAs are consistent with each other. 
    The errors are $1\sigma$ confidence levels.
    }
    \label{fig:PA_PD_ene}
\end{figure}



\begin{table*}
    \centering
    \caption{The result of the spectro-polarimetric fits for the time-averaged, eclipse, and non-eclipse intervals. The errors show $1\sigma =68.3\%$ confidence levels. 
    Note that the Fit stattis values ($\chi^2/\nu$) coressponds those of 2-8 keV spectro-polarimetric fits, respactively.
    }
    \begin{tabular}{c|c|c|c|c}
       models   & parameters  & values (Time averaged) & values (Non-Eclipse) & values (Eclipse) \\
       \hline
       {\tt TBabs}               & $N_H [10^{22} ~{\rm cm^{-2}}]$ & 0.15 (fix)  & 0.15 (fix)          & 0.15 (fix)\\
       {\tt polconst} (2-8 keV)  & PD [\%]                        & $8.5\pm1.6$          & $5.9\pm1.9$& $15\pm3$   \\
                                 & PA [deg]                       & $-46\pm6$            & $-52\pm10$& $-41\pm6$ \\ 
                                                                                                        
       {\tt polconst} (2-4 keV)  & PD [\%]                        & $8.4\pm2.0$          & $4.7\pm2.3$& $17\pm4 $ \\
                                 & PA [deg]                       & $-37\pm7$            & $-41\pm15$& $-37\pm6$ \\ 
                                                                                                        
       {\tt polconst} (4-8 keV)  & PD [\%]                        & $12\pm3$             & $11\pm4$  & $15\pm6$ \\  
                                 & PA [deg]                       & $-61\pm7$            & $-61\pm10$& $-53\pm11$ \\
         
       {\tt diskbb}              & $kT_{\rm in}$ [keV]            & $1.2\pm0.2$  & $1.1\pm0.2$ & $1.6^{+0.3}_{-0.5}$\\
       
                                 & norm     & $0.90^{+0.80}_{-0.30}$   & $1.28^{+0.82}_{-0.58}$ & $0.46^{+0.88}_{-0.16}$\\
       {\tt bbody}  & $kT$ [keV] & $2.5^{+0.7}_{-0.5}$  & $2.2^{+0.8}_{-0.3}$& $>2.6$\\
         & norm $[\times 10^{-4}]$ & $7.8\pm1.4$ & $7.4\pm0.9$& $15.0^{+9.8\times10^{3}}_{-8.6}$ \\
    ${\rm constant_{1}}$& $ C_{\rm DU1} $ & 1.0 (fix) & 1.0 (fix) & 1.0 (fix)\\
    ${\rm constant_{2}}$& $ C_{\rm DU2} $ & $1.02\pm0.01$  & $1.02\pm0.01$& $1.01\pm 0.02$\\
    ${\rm constant_{3}}$& $ C_{\rm DU3} $ & $0.98\pm0.01$  & $0.98\pm0.01$& $0.97\pm 0.02$\\
    \hline
    Flux (2-8keV) & $F_{\rm DU1} [\times 10^{-11} {\rm erg~cm^{-2}~s^{-1}}]$ & $4.83$  & $4.70$& $5.10$\\
    \hline
     Fit statistic &$\chi^2/\nu$ (2-8 keV: I, Q and U)& 1040/1090  & 987/1003 & 734/784\\
    \hline
    \end{tabular}
    \label{tab:time_average}
\end{table*}

We first perform simultaneous fitting of the time-averaged data sets from each detector unit (DU1, DU2, and DU3) using the model {\tt constant * TBabs * polconst * (diskbb + bbody)} in {\sc xspec}. This model assumes that the intrinsic spectrum comprises a low-temperature disc blackbody and a high-temperature blackbody originating from the neutron star boundary layer (Table~\ref{tab:time_average}). 
For the fitting procedure, we fix the Galactic absorption column density at $N_{\rm H}=0.15 \times 10^{22} {\rm cm^{-2}}$, consistent with previous analyses \citep{Tomaru2023b}. 
The {\tt polconstant} component in our model assumes a single polarization for the sum of  
the two intrinsic emission spectra, with constant 
PD and PA across the entire energy range. 
We incorporate cross-normalization factors between detector units using the {\tt constant} component. 

This model provides an excellent description of the spectro-polarimetric data with $\chi^2/\nu = 1040/1090$ (see the spectrum and Stokes Q and U parameters in Fig. \ref{fig:sp_pd}). 
It reveals a significant polarization signal with a PD of $8.5\pm1.6\%$ and a PA of $-46\pm 6^{\circ}$ (errors represent $1\sigma$ confidence levels) in the 2-8 keV band (Tab.\ref{tab:time_average}).
Its detection level is high at $\sim 4.8\sigma$  (one-sided Gaussian-equivalent significances relative to the null unpolarized hypothesis), which is calculated by the comparison of the fit statistic with the polarization component $\chi^2/\nu = 1040/1090)$ and without that ($\chi^2/\nu = 1068/1092$ with setting PD=0 in {\tt polconst}).

Fig. \ref{fig:pol_whole} (left) displays the two-dimensional confidence contours for PD and PA of this energy band (blue), clearly demonstrating the robust detection of both parameters.
We plot the time-averaged PD and PA as the orange band on the 3rd and 4th panels of Fig. \ref{fig:lc}, and also show the marginal variability about these mean values in the three time intervals. 

This PD is much larger than that measured for weakly magnetised neutron stars in ordinary LMXBs and provides strong evidence that the X‑ray emission from this source is produced by scattering in an accretion disc wind, as shown previously by \citet{Tomaru2023b}. This contrasts with PDs well below $5\%$ in typical (non ADC) LMXBs.
For example, the atoll source GX 9+9 shows $1.4\pm 0.2\%$ \citep{Farinelli2023}, 
and the more luminous Z-source Cyg X‑2 reaches only $1.8\pm 0.3\%$ \citep{Ursini2023}. 

The model above assumes only a single polarization, but the intrinsic spectrum is made from two components (disc and boundary layer) that should have different polarization properties \citep{Gnarini2022,Tomaru2024}.
Hence we explore whether there is any energy dependence of the polarization which might be associated with the switch between disc dominating at low energies, to the boundary layer dominating at higher energies.
We split the data into two energy bands, 
and show the resulting contours on 
Fig.\ref{fig:pol_whole} (left) for 
2-4~keV (orange) and 4-8~keV (green). 
There is a shift in best fit PD with energy, from 
$8.5-13\%$, and a shift in PA (from $-37^{\circ}$ to $-60^{\circ}$) but the contours clearly substantially overlap, so there is only marginal evidence ($1.5\sigma$) for energy dependence.

We repeat this analysis for the data split into non-eclipse (Fig.\ref{fig:pol_whole}: middle)
and eclipse (Fig.\ref{fig:pol_whole}: right)
time periods, but the shorter time segments mean that the uncertainties are larger and the significance of any energy shift between
2-4~keV (orange) and 4-8~keV (green) 
is less than $1\sigma$.

Nonetheless, spilting the data into just two energy bands only could be too crude to pick out a signal in noisy data. Instead, we return to the time averaged data and replace the {\tt polconst} model with {\tt pollin} in order model a linear dependence with energy so that 
${\rm PD}(E) = {\rm PD}_{1} + (E-1.0)\,{\rm PD}_{\rm slope}$ and
${\rm PA}(E) = {\rm PA}_{1} + (E-1.0)\,{\rm PA}_{\rm slope}$,
where $E$ is the photon energy in keV. 
This model provides a good description of the data (Tab.\ref{tab:pollin1}), with $\chi^{2}/\nu = 1032/1088$, and yields a modest improvement over the single {\tt polconst} model ($\Delta\chi^{2}=8$ for two additional degrees of freedom). 
The improvement is driven primarily by the energy dependence of PA: when ${\rm PD}_{\rm slope}$ is fixed to zero, the fit statistic becomes $\chi^{2}/\nu = 1033/1089$ (Tab.\ref{tab:pollin2}). 
Thus, in the time-averaged data, the evidence for an energy-dependent PA is stronger than that for an energy-dependent PD. However, there is no {\it a priori} reason to fix the PD and only allow the PA to vary, so an unbiased view of the statistical significance of the change in PA is still $\Delta \chi^2=8$ for 2 degrees of freedom. 
This is around $2.5\sigma$ significance. 
\begin{table}
    \centering
    \caption{The parameters taken by {\tt pollin} (Total spectral model is {\tt constant*TBabs*pollin*(diskbb+bbody)}).Note that other parameters are the same as Tab.\ref{tab:time_average}}
    \begin{tabular}{c|c|c|c}
    model & parameters & Time avareged &Eclipse\\
    \hline
       pollin  &  ${\rm PD_{1}} ~[{\rm \%}]$ & $<8.6$ &$18\pm8$\\
               &  ${\rm PD_{slope}} ~[{\rm \%}]$ & $2.0\pm 1.5 $ &$-0.5\pm2.9$\\
               & ${\rm PA_{1}~ [deg]}$ &$ -13\pm 13$&$-12\pm 14$\\
               & ${\rm PA_{slope}~ [deg]}$ &$ -11\pm 4$ &$-11^{+5}_{-6}$ \\
               \hline
        statistic & $\chi^2/\nu$ & 1032/1088 & 730/782\\
    \end{tabular}
    \label{tab:pollin1}
\end{table}

\begin{table}
    \centering
    \caption{As in Tab\ref{tab:pollin1}, but with $PD_{\rm slope}$ fixed to 0.}
    \begin{tabular}{c|c|c|c}
    model & parameters & Time avareged &Eclipse\\
    \hline
       pollin  &  ${\rm PD_{1}} ~[{\rm \%}]$ & $9.6\pm1.6$ &$17 \pm3$\\
               &  ${\rm PD_{slope}} ~[{\rm \%}]$ & $0$(fix) &$0$ (fix)\\
               & ${\rm PA_{1}~ [deg]}$ &$ -15\pm 13$&$-12\pm 14$\\
               & ${\rm PA_{slope}~ [deg]}$ &$ -11\pm 4$ &$-11\pm5$ \\
               \hline
        statistic & $\chi^2/\nu$ & 1033/1089 & 730/783\\
    \end{tabular}
    \label{tab:pollin2}
\end{table}

\begin{figure}
    \centering
    \includegraphics[width=0.9\hsize]{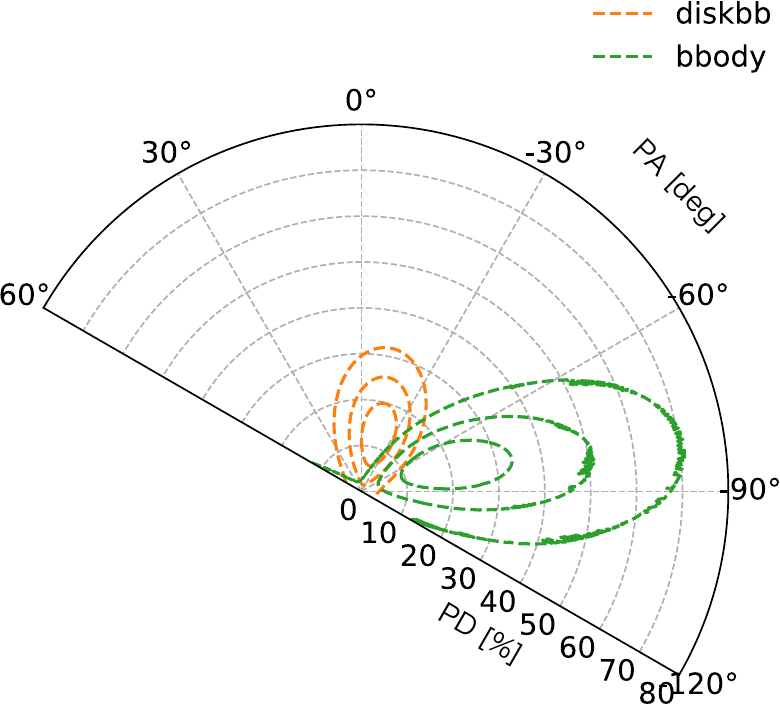}
    \caption{2-dimensional polarization contour plot for time-averaged data by 2 {\tt polconst} model($1\sigma, 2\sigma, 3\sigma$  confidence levels).
    The colours show {\tt diskbb} (orange) and {\tt bbody} (green).
    We extend the PA range from $-90^{\circ}$ to $90^{\circ}$ degrees to $-120^{\circ}$ to $60^{\circ}$ for better visualization.
    }
    \label{fig:pol_dbb_bbody}
\end{figure}

Physically, any change in PD and PA is likely associated with a difference in weighting of the two spectral components, so replace the {\sc pollin} with a more physically motivated model to investigate the potential energy dependence. We
fit the time-averaged data with 2 {\tt polconst} models, one which is applied to the 
disc spectrum, and a separate one for the boundary-layer emission. The results are tabulated in  Tab.\ref{tab:polconst_dbb_bbody}, and shown as a contour plot in Fig.\ref{fig:pol_dbb_bbody}. This shows evidence for a shift in PA of around $\sim 60^{\circ}$ at around the $2\sigma$ level.

We repeat this exploration of the energy dependence of the polarization on the eclipse data (see Tab.\ref{tab:pollin2} and Tab.\ref{tab:polconst_dbb_bbody}), but again find lower statistical significance for any change than for the time-averaged data.

\begin{table}
    \centering
    \caption{The parameters taken by 2 {\tt polconst} models.
    Note that other parameters arethe  same as Tab.\ref{tab:time_average}}
    \begin{tabular}{c|c|c|c}
    model & parameters & Time avareged &Eclipse\\
    \hline
       ${\tt polcosnt_{\rm disc}}$ &  ${\rm PD} ~[{\rm \%}]$ & $12^{+5}_{-4}$ &$19^{+10}_{-7} $\\
               & ${\rm PA~ [deg]}$ &$ -18^{+11}_{-10}$&$-23^{+13}_{-12}$\\
        ${\tt polconst_{\rm bbody}}$       &  ${\rm PD} ~[{\rm \%}]$ & $19^{+8}_{-6}$ &$28^{+34}_{-14}$\\
               & ${\rm PA~ [deg]}$ &$ -73\pm10$ &$-73^{+15}_{-17}$ \\
               \hline
        statistic & $\chi^2/\nu$ & 1033/1088 & 731/782\\
    \end{tabular}
    \label{tab:polconst_dbb_bbody}
\end{table}

In summary, the data include an eclipse but the statistics are not good enough to strongly constrain any difference between the polarisation in eclipse and out of eclipse, though there is weak evidence ($1.5\sigma$) for an increase in PD during eclipse. Hence we use the time averaged spectrum to maximise statistics. This shows some evidence ($\sim 2\sigma$) for a change in PA with energy, with weaker evidence for an increase in PD with energy.

\section{Modeling of polarization}
\begin{figure*}
    \centering
    \includegraphics[width=0.9\linewidth]{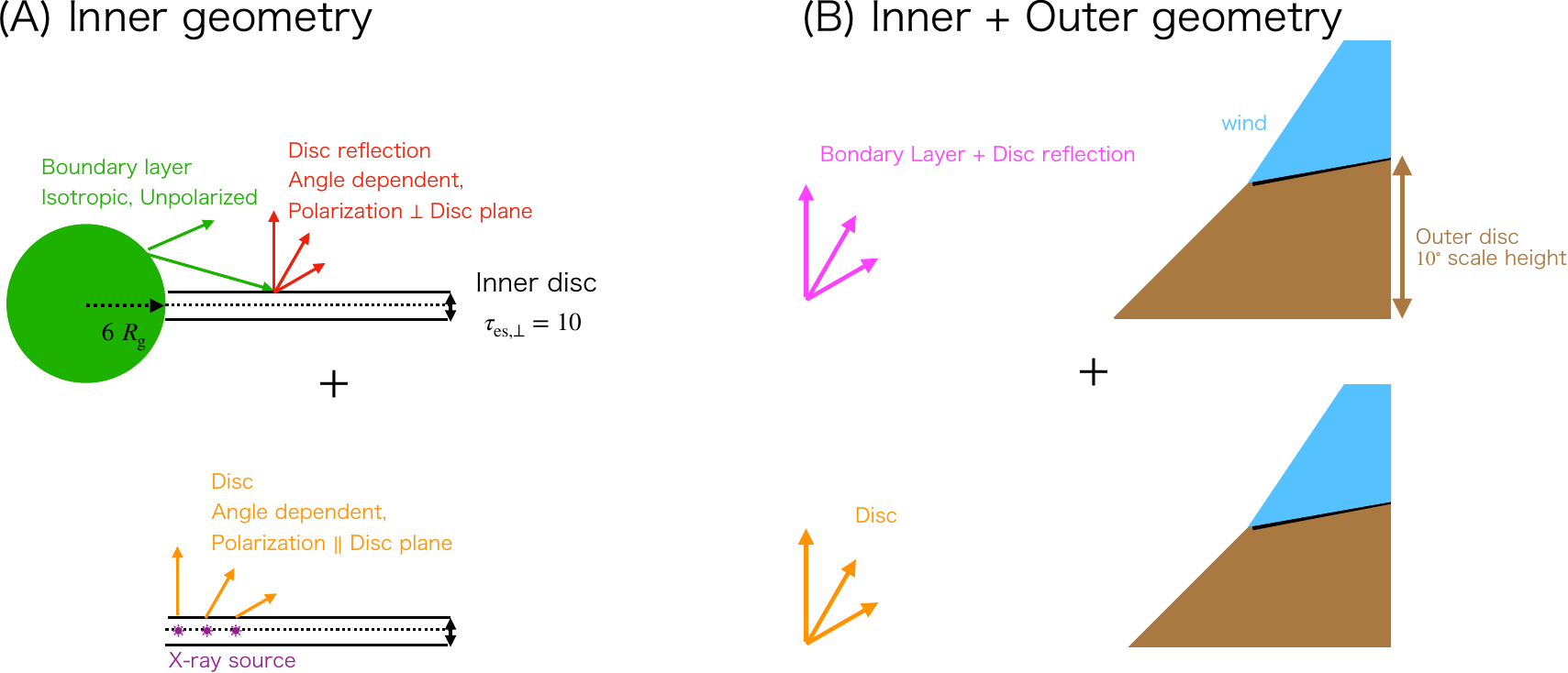}
    \caption{The schematic geometry of the model. (A) The model of the normal lower magnetised neutron star. We calculated different source distributions and combined them to predict total polarization. 
    (B) The model of the Accretion disc corona source. We add the outer disc and disc wind to the model (A).
    }
    \label{fig:model_geometry}
\end{figure*}

To understand the origin of the polarization from this source, we use the Monte-Carlo radiation transfer code {\sc monaco}  \citep{Odaka2011, Tomaru2024}. 
This code calculates the interaction between photons and electrons and tracks the propagation of each photon, including polarization from electron scattering. 
The effect of special relativity is also take into account.

We only consider Compton scattering (both up scattering and down scattering), assuming the material near the compact object is completely ionised. 
This is appropriate for the intrinsic emission components, but we note that the wind and the inner disc reflection are not completely ionised, so there can be additional absorption opacity which can slightly change the energy dependence \citep{Tomaru2024,Podgorny2024}.

We first calculate the polarization expected from the inner components : the boundary layer and its inner disc reflection (top left in Fig.\ref{fig:model_geometry}), plus intrinsic disc emission (bottom left in Fig.\ref{fig:model_geometry}). 
This should give results which are appropriate for short period NS LMXRB, where the outer disc is too small to launch a thermal-radiative disc wind \citep{Done2018}.
We then include the effect of scattering of these components in an outer disc wind (geometry B in Fig.\ref{fig:model_geometry}).
This should be appropriate for long period NS LMXRB systems, and we explicitly calculate extremely high inclination angles where the disc obscures the central source so as to be able to model the ADC sources.

In this calculation, we assume an axisymmetric geometry. Although the source–disc configuration is axisymmetric, the relativistic Keplerian rotation of the disc introduces Doppler boosting and aberration. These special-relativistic effects break the fore–aft symmetry between the approaching and receding sides of the disc in the observer’s frame, so that the azimuthal cancellation of the Stokes parameter $U$ is no longer exact. As a result, we obtain $U\ne0$ and a polarization position angle that is slightly rotated away from the symmetry axis.

We define the direction of the disc rotation axis as PA$=0^{\circ}$, and the Stokes parameter $Q$ is positive when the polarization direction is perpendicular to the disc plane (i.e., aligned with the disc rotation axis), whereas it is negative when the polarization direction is parallel to the disc plane.
We note that the Stokes parameter $U$ is generally non-zero in our calculations due to the relativistic effects, but its value is always smaller than $Q$ (i.e., $|U|\leq |Q|$), so the polarization position angle remains close to either $0^{\circ}$ or $\pm 90^{\circ}$.


\subsection{Modeling short period NS LMXBs}
\label{sec:normal_NS_LMXB}
\begin{figure}
    \centering
    \includegraphics[width=0.9\linewidth]{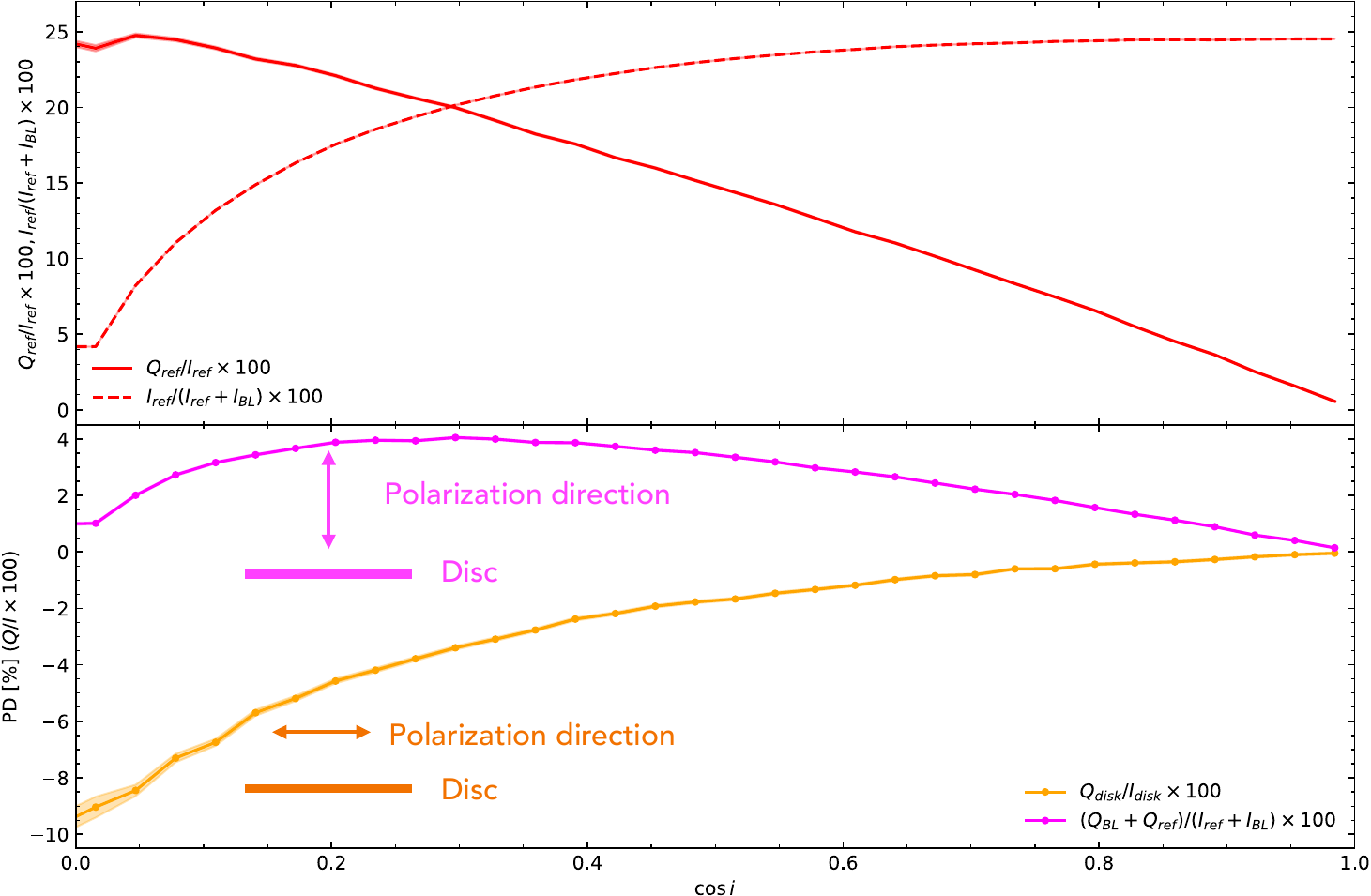}
    \caption{The energy integrated (2-8 keV) polarization properties of all the model components for geometry A.
    Upper panel: The reflected spectrum PD as a function of inclination (red).
    The dashed line shows the fraction of reflection to the total boundary layer plus reflected emission.
    Lower panel: Multiplying the fraction of reflection and its PD together gives the PD of the boundary layer plus its reflection from the disc (magenta line). 
    The orange line instead shows the orthogonally polarized intrinsic disc emission.
    }
    \label{fig:PD_NS_disc}
\end{figure}

For the first simulation, we calculate the polarization 
expected from all the components of a 
(low-magnetisation) NS LMXRB (A in Fig.\ref{fig:model_geometry}). 
We set a spherical source geometry with a radius of $6 R_{g}$ to mimic the emission from a boundary layer which covers the entire NS surface, so it is spherical (isotropic) and unpolarized (see \citealt{Bobrikova2025} for a discussion of other boundary layer geometries). 
Guided by the observed spectrum, we assume this emits like a black body with $kT_{\rm bb}= 2.2~{\rm keV}$.

We follow $7.6\times 10^8$ intrinsically unpolarized photons from this boundary layer, allowing them to reflect from a completely ionised, optically thick, geometrically thin disc. 
We model this as extending from 
$ R_{\rm in}\leq R \leq R_{\rm out}$, where $R_{\rm in}=6 R_{g}$ and $R_{\rm out} = 600 R_g$.
We make a logarithmic radial grid as $R_{i} =R_{\rm in}(R_{\rm out}/R_{\rm in})^{i/N_R}$ ($i = 0, 1...N_{R}$, $N_R=32$).
We set temperature distribution as $T_{\rm disc}(R)=T_{\rm in}(R/R_{\rm in})^{-3/4}$ and $kT_{\rm in} = 1.1 {\rm keV}$, which is the observational value taken by Non-Eclipse value of {\ixpe} (Tab.\ref{tab:time_average}).
Using the scale height of this temperature distribution, we assume the density of disc as $n_{e}(R) = \tau_{\rm es}/(\sigma_{\rm es} H)$, where scale height as $H = (c_s/\Omega_{k})$, $c_s  =\sqrt{kT_{\rm disc}(R)/(\mu m_p)}$ and $\Omega_k = \sqrt{GM/R^3}$. 
We assume the vertical optical depth is 5 from the mid-plane to the disc surface, which means the total vertical optical depth is $\tau_{\rm es} =10$.
We also add the Keplerian rotation velocity to the disc.

The upper panel of Fig.\ref{fig:PD_NS_disc} shows  the inclination angle dependence of the reflected PD (2-8 keV band, red).
This component is polarized perpendicular to the disc plane (we define this as PA$=0^{\circ}$, equivalently $Q>0$), with the PD increasing towards more edge-on viewing angles.
The 2--8~keV band-integrated PA is consistent with $0^{\circ}$ (i.e. $U\simeq 0$), because Doppler boosting is not sufficiently strong in this energy range).
The reflected intensity also has an angle dependence, similar to that for the intrinsic disc spectrum in that it is brighter for more face on geometries. 
The dashed red line in the upper panel of Fig.\ref{fig:PD_NS_disc} shows the fraction of reflected emission to the total (boundary layer plus reflected) emission. The fraction is large for face on geometries (up to 25\%) but its polarization here is small, $\approx 0$, as a face on disc is circular.
Conversely, the reflected emission makes only a small contribution to the total spectrum at high inclination (less than 10\%), but it is highly polarized at $\approx 25$\%.
This means that the combination of boundary layer plus reflection (purple, lower panel of Fig.\ref{fig:PD_NS_disc}) 
which is simply given by multiplying the reflected polarization by the fraction of reflection in the total spectrum, produces a total polarization which has a maximum around PD=3--4\% for angles from $60-80^\circ$, and 2--3\% for $30-60^\circ$.
This result is consistent with the earlier work \citep{Lapidus1985}.

\begin{figure*}
    \centering
    \includegraphics[width=0.9\linewidth]{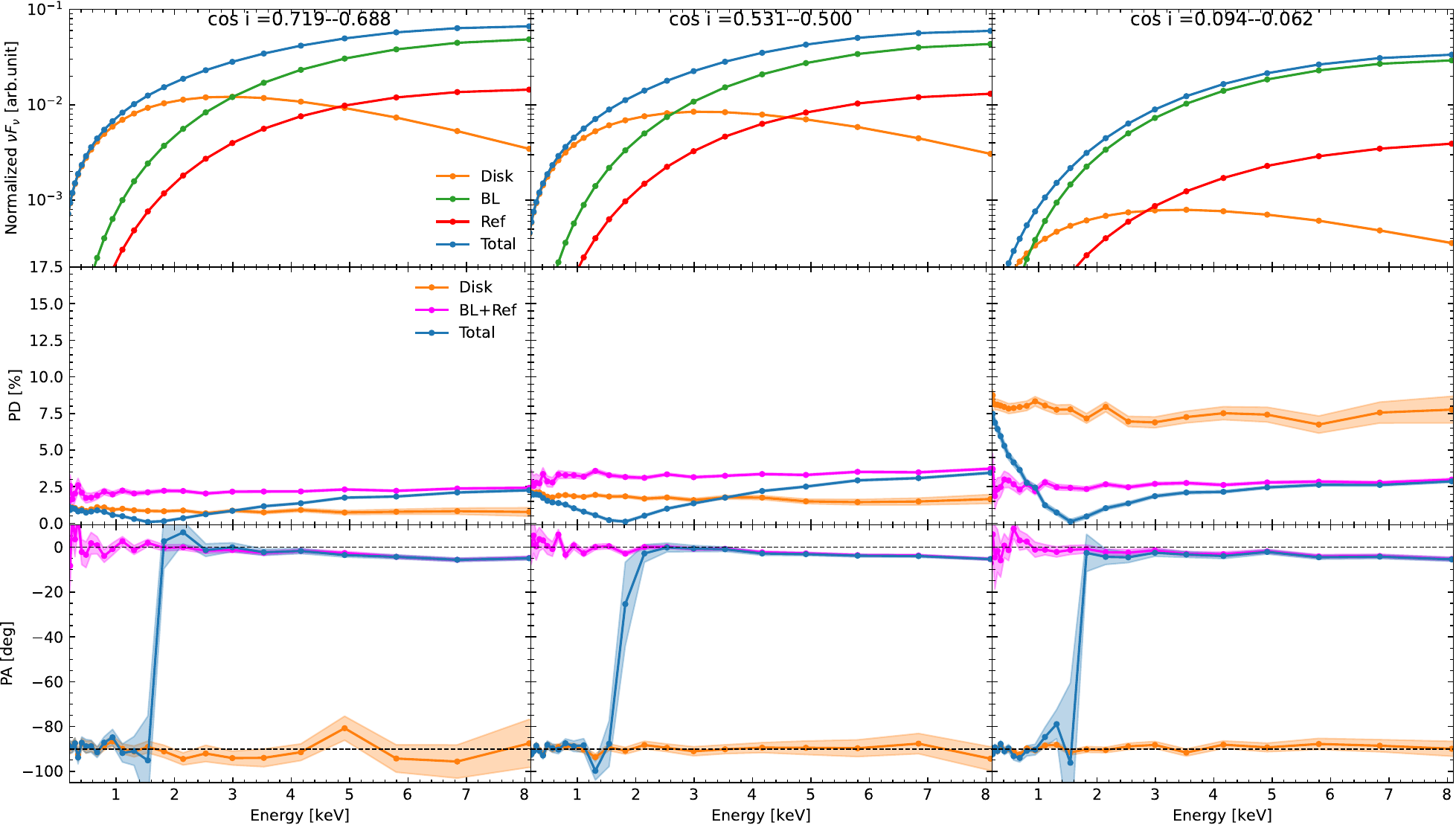}
    \caption{The spectra-polarimetric model for low mass X-ray binaries with three different inclination angle bins. 
    The top panels show the spectrum of each component. The boundary layer (green) is mostly isotropic, but its reflection (red) and the intrinsic disc emission (orange) are relatively brighter face on. 
    The blue line shows the total spectrum.
    The middle panels show the PD with energy for each component.
    Since the boundary layer and its reflection have the same spectrum (see upper panel), their total emission is co-added (magenta) to show the PD of this combined component. 
    This is polarized perpendicular to the disc, so has ${\rm PA}\sim 0$ while the intrinsic disc emission (orange) is polarized parallel to the disc so has ${\rm PA}\sim \pm90$.
    The total spectrum (blue) switches from being dominated by the disc at low energies to being dominated by the boundary layer and its reflection at high energies, so there is an increase in PD with energy, leading also to a swing in PA as shown in the lower panel.
    We note that the non zero PA at higher energy is seen due to the Doppler beaming. 
    }
    \label{fig:nufnu_QI}
\end{figure*}

Our calculation assumes that the reflection/scattering occurs in fully ionised material, and it is reasonable near the compact object due to strong photoionization.
In this limit, the  PD is nearly energy independent (cyan curve in Fig.~\ref{fig:nufnu_QI}), while the PA shows only a small variation, likely induced by relativistic beaming.
A more realistic reflector is expected to be partially ionized.
As demonstrated by \citet{Podgorny2025}, partially ionized material can produce an energy-dependent PD because photoelectric absorption preferentially removes photons that would otherwise undergo multiple scatterings within the slab/disc atmosphere, especially in the low ionization state.
By suppressing multiple-scattering contributions, the emergent polarized signal becomes more sensitive to energy, leading to structured PD(E) behavior.
This could also produce energy-dependent PD, as we observed several sources \citep{Ursini2024}. 

We next distribute the source photon within the inner disc mid-plane following the temperature distribution of the disc.
Thus the photon distribution integral with energy is $n_{\rm ph}(R) [{\rm photons~s^{-1}~cm^{-2}}]\propto T_{\rm disc}(R)^3 =R^{-9/4}$.
We generate the same number of photons as the boundary layer ($7.6\times 10^8$ photons), and these electron scatter out of the disc to produce the standard Chandrasekhar polarization and intensity pattern \citep{Chandrasekhar1960}.
The polarization direction is parallel to the disc plane (PA=90, equivalently PD<0), and the maximum PD is about $11\%$ at $\cos i=0$. 
This result is consistent with that of the previous work \citep{Tomaru2024}.
We show this compared to the boundary layer plus reflected emission as the orange line on the lower  panel of Fig.\ref{fig:PD_NS_disc}.

As well as different angle dependences, these components have different energy dependences.
We assume here that the source has intrinsically similar intensity from both the disc and boundary layer (as is expected in Newtonian gravity, where half of the available gravitational potential energy is radiated in the disc, and the other half kept as kinetic energy in rotation which can be released in the boundary layer)  i.e. that 
$\int \int E n_{\rm ph0,disc} (E, \Omega) dE d\Omega =\int \int E n_{\rm ph0,BL}(E, \Omega) dE d\Omega$, where $n_{\rm phi0, disc}$ is the number of photons per energy per solid angle from the boundary layer, and $n_{\rm ph0, disc}$ is the same for disc. 

The top panel in Fig.\ref{fig:nufnu_QI} shows the spectrum resulting from this at three different inclination angles ($i=45, 60, 85^{\circ}$), with the disc (orange), boundary layer (green), reflection (red) and total (blue).
The disc and reflected spectrum both drop with angle, while the boundary layer remains constant.
Our assumption of a completely ionized disc means that the reflected spectrum is identical in its energy dependence to that of the boundary layer, so we add these two together to consider their polarization properties. 
The middle panel shows the PD as a function of energy for the boundary layer plus reflection (magenta) and the disc (orange), together with the total spectrum (blue). 
There is now a distinct increase in PD for the total spectrum as a function of energies due to the spectral change from being dominated by the disc at low energy to being dominated by the boundary layer and its reflection at high energy.
The bottom panel shows the total PA as a function of energy. This shows a distinct swing from low energies, where the disc dominates and is intrinsically polarized parallel to the disc (${\rm PA} =-90^{\circ}$) to high energy where the boundary layer and its reflected photons dominate, which is polarized perpendicular to the disc (${\rm PA}=0^{\circ}$) around 2 keV. 
The higher the energy band, the more the effect of Doppler beaming appears, which breaks the symmetry and gives ${\rm PA}\neq0$, since the higher energy band photons tend to come from the inner fast rotation region near the NS. However, the effect of this is quite small, with ${\rm PA}\sim -5^\circ$.

This polarization model can explain the PD found by {\ixpe} in some low magnetised NS in LMXBs, such as  GX 9+9 (orbital period of 4.19 hours), where the PD  slightly increases with energy from  $1.4\pm0.4 \%$ (2-4 keV) to $2.2\pm5 \%$ (4- 8 keV) \citep{Ursini2023}. \citet{Ursini2023} explicitly fit reflection using the best current models i.e. including the residual atomic opacity of the highly ionised disc. This gives a good fit to the spectrum, but their reflection normalisation is smaller than predicted by our simulations here for a low-to-moderate inclination, so they require a much higher polarization of their inferred reflected emission. Reflection fitting is complex for highly ionised hot discs, and it may be that they are not sensitive to a completely ionised reflection component from the very innermost disc, and hence they are underestimating the amount of reflection present, and so overestimating the polarization required for their observed reflected component.

\begin{figure}
    \centering
    \includegraphics[width=0.9\hsize]{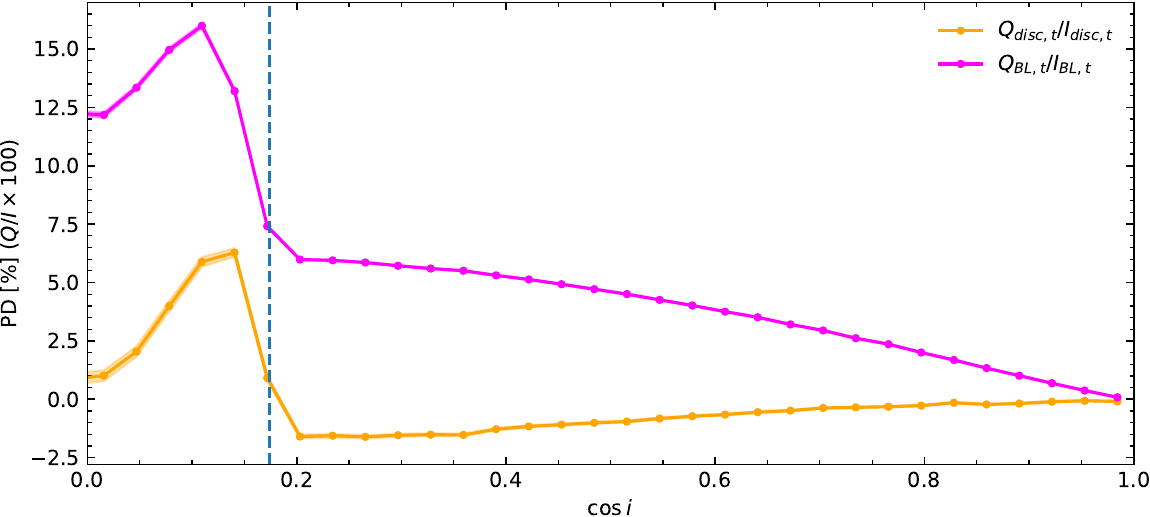}
    \caption{The energy integrated (2-8 keV) polarization properties of all the model components for geometry B (including the outer disc and wind). 
    The boundary layer and its reflection (magneta) have around 1\% more polarization perpendicular to the disc than for the equivalent non-wind (geometry A) simulation. Similarly, the intrinsic disc emission (orange) exhibits additional polarization from the wind, but because it is orthogonal, it leads to a 1\% decrease in the disc's PD. 
    The dashed line at $\cos i=0.17$ is the inclination of $80^\circ$ where the disc blocks a direct view of photons from the central source.
    The observed emission is then dominated by wind scattered photons, so again this adds to the PD of the boundary layer and its reflection, but is orthogonal, so it suppresses the scattered polarization of the disc photons. 
    }
    \label{fig:wind_pol}
\end{figure}

\begin{figure*}
    \centering
    \includegraphics[width=0.9\linewidth]{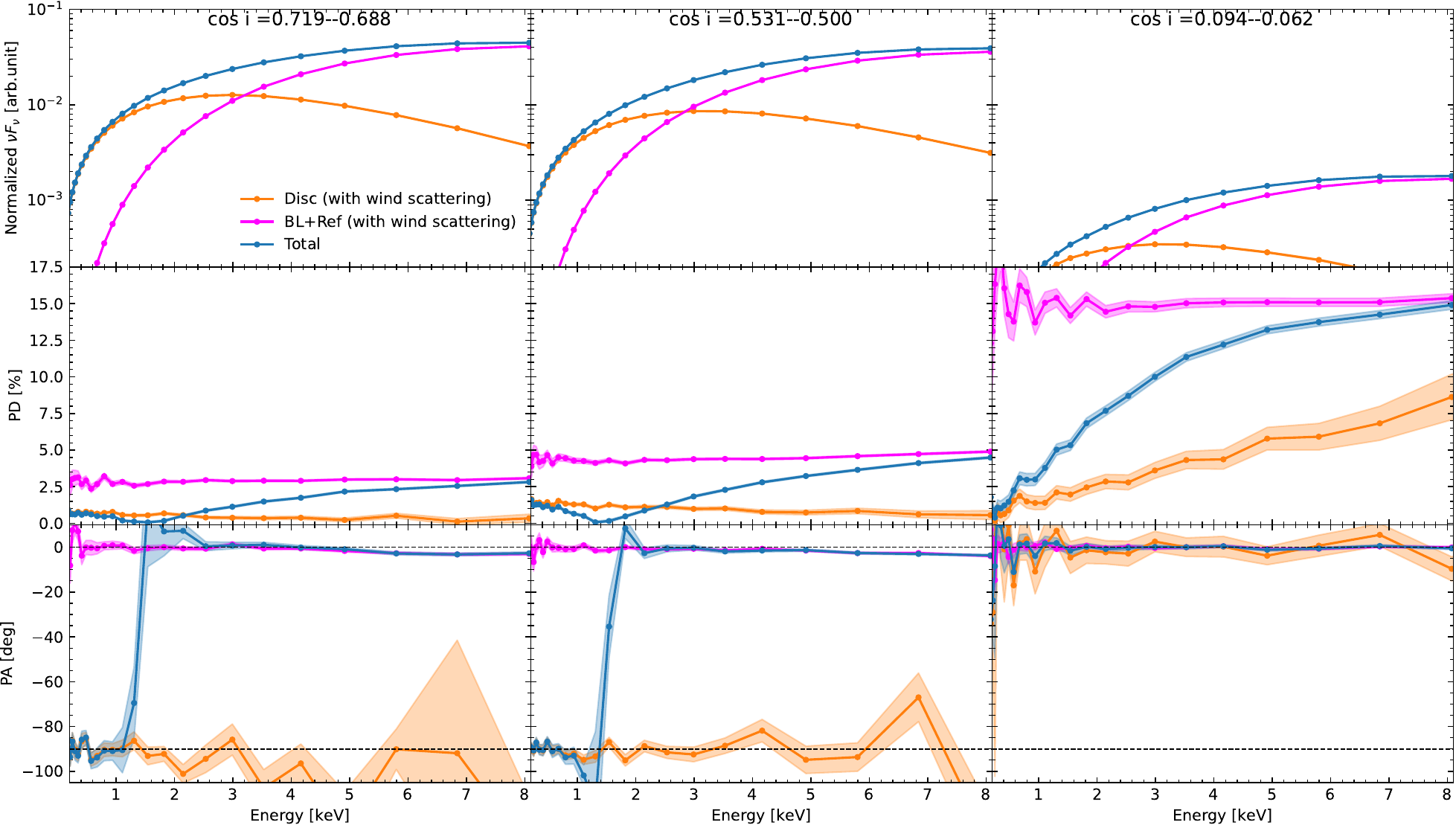}
    \caption{As Fig.\ref{fig:nufnu_QI} but with the disc wind. This makes only a small difference in PD at low-to moderate inclinations, but the disc can now completely block the direct emission from the central source at high inclinations, so that wind scattering dominates the flux. }
    \label{fig:nufnu_QI_wind}
\end{figure*}

\begin{figure}
    \centering
    \includegraphics[width=0.9\hsize]{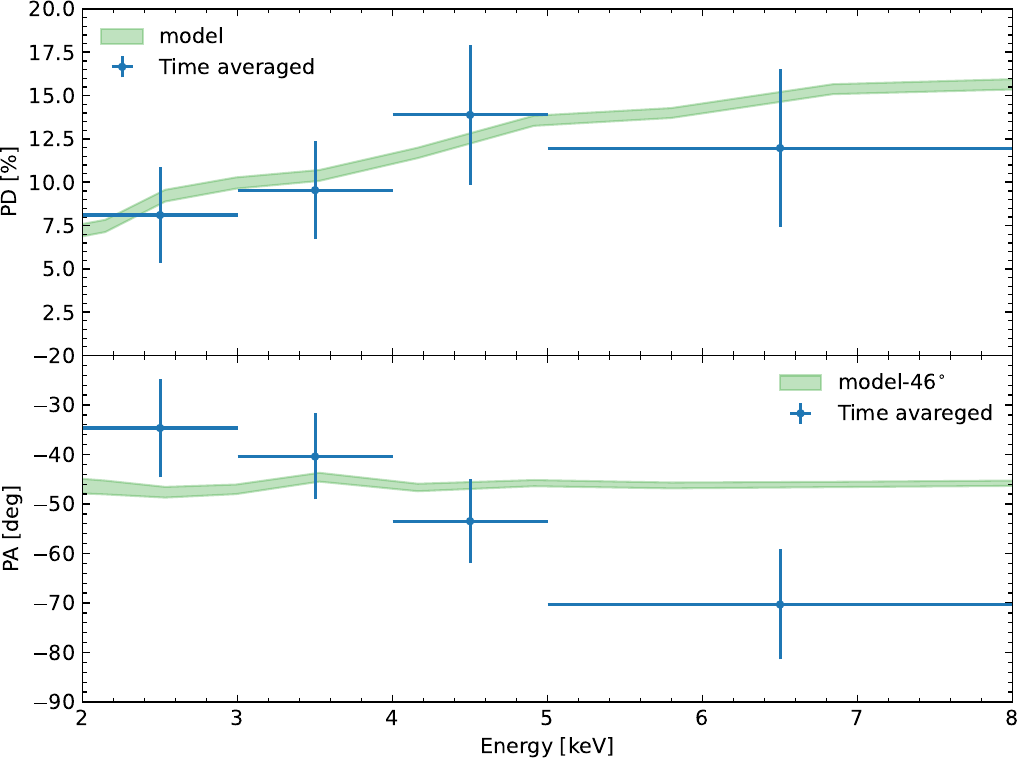}
    \caption{The comparison between the time-averaged energy-dependent PD/PA in the data (blue points) and the model including wind scattering (green band).
    The model can reproduce energy denedence of the PD, but can not reproduce that of the PA (Fig.\ref{fig:PA_PD_ene}).
    }
    \label{fig:model_data}
\end{figure}

\subsection{Modelling long period NS LMXRB}
\label{sec:ADC_model}

A long period, bright NS LMXRB should also have substantial thermal-radiative winds from X-ray illumination of the outer disc.
Thus, we add the wind structure outside of the previous disc (see B in Fig.\ref{fig:model_geometry}) to predict the additional impact on the polarization properties imprinted by scattering in the wind. 

As a simulation setup, we set the same column density structure $\log N_{H} = 22+2.4(1-\cos i), ( 0\leq \cos i \leq 0.174)$ as \citep{Tomaru2024}, although a different column density distribution will change the predicted polarization \citep{Nitindala2025}.
The maximum wind inclination angle 
is set by our assumed boundary of the wind and the outer disc at
$i_{\rm w}=80^\circ$ since the inclination of this object is $i\sim 83^{\circ}$ \citep{Ashcraft2012}.
At a higher inclination angle than $i=80^\circ$, we set an optically thick disc to separate each wind region as $\tau_{\rm z} = 5 $ from the mid-plane. 
This gives the radial optical depth for the outer disc as $\tau_{R}\sim \tau_{\rm z} /\tan(10^\circ)\sim 28$.  

This outer wind can polarize the scattered X-rays by $10-20 \%$, in a direction which is perpendicular to the disc 
\citep{Tomaru2024}.
Isotropic illumination gives a larger PD from the wind scattered flux ($\sim 20\%$) than the disc ($\sim 10\%$), as the former illuminates the more equatorial region, where the polarization perpendicular to the disc is larger (see also \citealt{Nitindala2025}).
Thus, scattering in the wind will give larger polarization to the NS boundary layer than the disc (and reflected emission).

Fig.\ref{fig:wind_pol} shows the inclination dependence of the energy-integrated PD including the wind scattering component on each of the boundary layer and its reflected emission (magenta) and the disc (orange).
Compared to the lower panel of Fig.\ref{fig:PD_NS_disc}, the boundary layer PD increases by 1-2\% where the source is seen directly as the additional scattering in the wind gives polarization in the same direction as the reflection of the boundary layer from the disc.
Conversely, the PD of the disc decreases by $\sim 1-2$\% as the intrinsic disc emission is polarized orthoganlly to the scattered flux.
This small change of polarization is consistent with the previous simulation \citep{Tomaru2024} at $i<80$, where the vertical disc blocks the entire direct source emission.
At the highest inclinations, $i>80^\circ$ ( $\cos i <0.17$), the direct photons from the inner region are completely blocked by the outer disc, so that the observed emission is dominated by scattering in the wind and the PD rises sharply, as the wind scattered photons have high polarization perpendicular to the disc.

Fig.\ref{fig:nufnu_QI_wind} is the same as 
Fig.\ref{fig:nufnu_QI} but with the additional outer 
wind structure which can change the observed spectral and polarization properties. 
For low to moderate inclination systems ($i=45, 60^\circ$), scattering by the wind imprints an additional PD of around $1\%$ in the direction perpendicular to the disc plane. The effect is small, as the ratio of wind scattered photons to the direct photons from the inner region is small. 
At the higher inclination than the outer disc ($\cos i <0.17$), the photons from the inner geometry are completely blocked (see B in Fig.\ref{fig:model_geometry}), but scattered in the outer wind.
This effect produces very high polarization, with PD of up to $\sim 15\%$ at 8 keV.
The total spectrum is dominated by the scattered boundary layer, as the reflected and intrinsic disc photons are suppressed at the high inclination angles, which are most effectively scattered in the wind. 
However, there is scattered disc emission at low energies, and this has quite low PD as it scatters on average higher in the wind, where there is more circular symmetry.
Thus, the PD exhibits a strong energy dependence. Still, the PA is constant, even when the disc component dominates at low energies, because it is almost entirely from scattered boundary-layer photons. These all have direction perpendicular to the disc (${\rm PA}=0^{\circ}$).
We note that there is no longer any significant energy dependence of the PA, because Doppler beaming is not directly observed: at these high inclinations, the observed flux is dominated by photons scattered in the outer disc wind.

\section{Match to the ADC source data and discussion}
\label{sec:discussion}

We now compare these model results with the time-averaged \ixpe\ data for the ADC source 2S~0921--630 (Fig.~\ref{fig:model_data}). 
We use the time-averaged dataset because the polarization measurement outside the eclipse is not statistically significant.
We note, however, that the time-averaged data include the eclipse interval, which is not incorporated in our model.
For a direct comparison, we rescale the model components (boundary-layer emission plus its reflection, and the disc emission) so that their sum reproduces the shape of the observed time-averaged 2--8~keV spectrum (blue curve in Fig.~\ref{fig:nufnu_QI_wind}) at an inclination of $i=83^{\circ}$.
The resulting model provides a good match to the observed PD, given the current uncertainties (top panel of Fig.~\ref{fig:model_data}). However, the model predicts a constant PA across the \ixpe\ bandpass, and therefore cannot reproduce the observed (and more significant) energy-dependent rotation of the $\Delta {\rm PA}\sim 40-60^\circ$.

It is very difficult to explain an energy-dependent rotation of PA which is {\em not} $\Delta {\rm PA}\sim 90^\circ$. Any axisymmetric geometry naturally gives polarization either aligned with the disc or aligned perpendicular to the disc. Our model includes the special relativistic Doppler boosting effect which does break the axisymmetry of the intrinsic emission, but this is a small effect, giving only $\Delta {\rm PA}\sim 5^\circ$ (see Fig:\ref{fig:nufnu_QI})
and this small signal on the boundary layer emission is wiped out when wind scattering dominates (see Fig:\ref{fig:nufnu_QI_wind}).

Instead, there is an obvious non-axisymmetry in our data as we sample an eclipse during the {\ixpe} exposure, where the companion star progressively blocks different azimuths of the scattering wind. Nonetheless, an eclipse by just the companion star is itself quite symmetric, and our data sample all of the eclipse, so any non-axisymmetries produced during the start of the eclipse when the approaching side of the disc-wind is occulted should
be equal and opposite to those at the end of the eclipse when the occultation switches to the receding side. 
This is consistent with the fact that the PA–energy trend measured from the time-averaged data and that extracted during eclipse are mutually consistent within the statistical uncertainties.

However, there are other potential non axisymmetries in the system geometry. Some possibilities are connected to the companion star. Firstly, this probably also has a wind, as seen in the ingress/egress eclipse timing in other binary systems
\citep{Knight23}. There is also a bulge on the outer disc where the accretion stream impacts \citep{Hellier1989} which can affect the polarization \citep{DiMarco2025}.
Other possibilities are connected to disc tilting between the neutron-star spin axis and the angular momentum axis of the accretion disc.
If the inner accretion flow/boundary layer is referenced to the stellar spin axis, while the outer disc and wind are referenced to the disc rotation axis, then the two polarization components can acquire different preferred orientations that are neither strictly parallel nor strictly perpendicular in the observer’s frame.
In addition, when the disc expands to large radii as this object, radiation forces can induce warping of the outer disc, 
and stably (or chaotically) precess via interaction with the tidal torques \citep{Ogilvie2001}, altering the geometry of both illumination and scattering.
A warped or tilted outer disc can in turn modify the wind launching geometry and introduce azimuthal asymmetries, shifting the net PA of the scattered component.

In this framework, the observed PA offset could be caused by the compounded effect of (i) component mixing during eclipse, and (ii) moderate misalignments among the relevant symmetry axes (spin axis, inner-flow axis, outer-disc axis, and wind geometry).
However, all these possible extensions to the geometry to make it asymmetric will only give a swing in PA with energy in a scattering dominated ADC source if the wind scattered disc component (low energy) has an intrinsically different PA to the wind scattered boundary layer (higher energies).
The more polar focussed emission pattern of the disc means that it preferentially scatters higher in the wind than the more isotropic boundary layer, but there is not much difference in scale height in our, quite equatorial wind model \citep{Tomaru2024}.
Instead, a swing in PA from the scattered disc component to the scattered boundary layer is more likely with a larger scale height wind, where there is a larger difference in mean height between the disc and boundary layer scattering regions \citep{Nitindala2025}.

\section{Summary}

We detect highly polarized X-ray flux from 2S 0921-630 with time-averaged PD $=8.5\pm 1.6 \%$ across the full 2-8~keV bandpass. 
The high polarization shows that the observed X-rays are scattered photons, confirming that the source is indeed an ADC, with the direct source emission blocked by the very high orbital inclination.
The data include an eclipse, and we find weak evidence ($1\sigma$) for an increase in PD to $15\pm 3$\% during the eclipse, compared to out of eclipse 
$({\rm PD} = 5.9 \pm 1.9\%$), with no change in PA. 
The energy-resolved analysis of the time averaged data shows a hint of increasing PD with energy and a more significant ($2\sigma$) swing of $\Delta PA\sim 40-60$ with energy.

We develop a full spectro-polarimetric model for short period NS LMXRB in general, as well as our ADC source in particular. We assume that the intrinsic source emission is the sum of an isotropic, non-polarized boundary layer, its reflection (angle dependent, similar to this, polarized perpendicular to the disc) and the 
disc itself (angle dependent, polarized parallel to the disc). We assume that the integrated (over energy and solid angle) luminosities from both disc and boundary layer are the same, as appropriate for a bright atoll system, and this gives up to 3--4\% polarization 
perpendicular to the disc from the contribution of the 
reflected emission for angles $i> 60^\circ$. 
This should be appropriate for short period systems, where 
the disc is too small for a substantial thermal-
radiative wind to form from the outer disc e.g. GX 9+9 
\citep{Ursini2023}.

We also extend the models to long-period, bright systems where there can be a substantial outer disc wind which imprints additional polarization from scattering (with direction perpendicular to the disc).
This generally increases the PD of the total spectrum by around 1\% 
as it adds to the PD from the reflected boundary layer contribution. This compares well with the X-ray polarization seen in the bright NS LMXRB systems such as Cyg X-2, GX 5-1 and other Z sources \citep{Gnarini2025}.

This extended model at very high inclination angles, where the direct photons are blocked by the outer disc, produces very high polarization. 
This can quantitatively match the PD seen in our data from the ADC source 2S 0921–630, and the model predicts an increase in PD with energy, which is consistent with the observed $1\sigma$ significance trend in the time -averaged data. 
However, our model cannot reproduce the $2\sigma$ significance swing in PA with energy. If confirmed with better data, this would show that there is an additional non-axisymmetric geometry which is not included in our axisymmetric model. 

This could be connected to the
companion star, either directly to the eclipse itself, especially if the star is being ablated by wind from the irradiating X-rays
\citep{Knight2023}. 
Alternatively it could be from the bulge resulting from the impact of the accretion stream onto the disc (e.g. \cite{Hellier1989}), and this would also make the wind from the outer disc somewhat non-axisymmetic also. Instead, the non-axisymmetry could be connected to the disc itself, as large discs are subject to radiation driven warping and this, coupled with the tidal torques, can lead to outer disc precession. This adds to the growing evidence for non-axisymmetric structures in NS LMXRB from polarization studies \citep{DiMarco2025, Kashyap2025}.



\section*{Acknowledgements}

This work was supported by Grant-in-Aid for JSPS Fellows JP	24KJ0152 (RT), 22K18277, and
22H00128 (HO).
Numerical computations were in part carried out on Cray XD2000 at Center for Computational Astrophysics (CfCA), National Astronomical Observatory of Japan (NAOJ).
Numerical analyses were in part carried out on analysis servers at CfCA, NAOJ.
CD acknowledges support from STFC through grant ST/T000244/1 and Kavli IPMU, University of Tokyo. Kavli IPMU was established by World Premier International Research Center Initiative (WPI), MEXT, Japan. 


\section*{Data Availability}

 The {\ixpe } data are publicly available. 
 Access to the radiation transfer code is available on reasonable request from H.O.(odaka@ess.sci.osaka-u.ac.jp).



\bibliographystyle{mnras}
\bibliography{library} 





\bsp	
\label{lastpage}
\end{document}